\documentclass[journal]{IEEEtran}

\usepackage{cite}

\usepackage{graphicx}

\newcommand{\cm}[1]{\ensuremath{ {{\rm cm}^{#1}}}}
\newcommand{\ntp}[2]{\ensuremath{#1\times10^{#2} } }
\newcommand{\asb}[2]{\ensuremath{#1_{\rm #2} }}

\long\def\@makecaption#1#2{%
  \vskip\abovecaptionskip
  \sbox\@tempboxa{#1: #2}%
  \ifdim \wd\@tempboxa >\hsize
    #1: #2\par
  \else
    \global \@minipagefalse
    \hb@xt@\hsize{\box\@tempboxa\hfil}%
  \fi
  \vskip\belowcaptionskip}
\makeatother

\begin{document}
\begin{titlepage}

\thispagestyle{empty}
\def\thefootnote{\fnsymbol{footnote}}       

\begin{center}
\mbox{ }

\end{center}
\begin{flushright}
\Large
\mbox{\hspace{10.2cm} physics/0611272} \\
\end{flushright}
\begin{center}
\vskip 1.0cm
{\Huge\bf
Radiation Hardness of CCD Vertex Detectors
}
\vspace{2mm}

{\Huge\bf
for the ILC
}
\vskip 1cm
{\LARGE\bf 
Andr\'e Sopczak$^1$,
Khaled Bekhouche$^2$, 
Chris Bowdery$^1$, \\
Chris Damerell$^3$, 
Gavin Davies$^1$, 
Lakhdar Dehimi$^2$,\\ 
Tim Greenshaw$^4$, 
Michal Koziel$^1$,
Konstantin Stefanov$^3$,\\ 
Tim Woolliscroft$^4$, 
Steve Worm$^3$}

\smallskip

\Large 
$^1$Lancaster University, U.K.\\
$^2$LMSM Laboratory Biskra University, Algeria\\
$^3$CCLRC Rutherford Appleton Laboratory (RAL), U.K.\\
$^4$Liverpool University, U.K.

\vskip 2.5cm
\centerline{\Large \bf Abstract}
\end{center}

\vskip 2.cm
\hspace*{-0.5cm}
\begin{picture}(0.001,0.001)(0,0)
\put(,0){
\begin{minipage}{\textwidth}
\Large
\renewcommand{\baselinestretch} {1.2}
Results of detailed simulations of the charge transfer inefficiency
of a prototype CCD chip are reported. The effect of radiation damage
in a particle detector operating at a future accelerator is studied
by examining two electron trap levels, 0.17\,eV and 0.44\,eV below
the bottom of the conduction band. Good agreement is found between
simulations using the ISE-TCAD DESSIS program and an analytical
model for the 0.17\,eV level. Optimum operation is predicted to be
at about 250\,K where the effect of the traps is minimal which is
approximately independent of readout frequency. This work has been
carried out within
 the Linear Collider Flavour Identification
(LCFI) collaboration  in the context of the International Linear
Collider (ILC) project.
\renewcommand{\baselinestretch} {1.}

\normalsize
\vspace{3.5cm}
\begin{center}
{\sl \large
Presented at the IEEE 2006 Nuclear Science Symposium, San Diego, USA, and the \\
10th Topical Seminar on Innovative Particle and Radiation Detectors
(IPRD06), 2006, Siena, Italy, \\
to be published in the proceedings.
\vspace{-6cm}
}
\end{center}
\end{minipage}
}
\end{picture}
\vfill

\end{titlepage}

\newpage
\thispagestyle{empty}
\mbox{ }
\newpage
\setcounter{page}{0}

\title{Radiation Hardness of CCD Vertex Detectors\\for the ILC}

\author{\authorblockN{
Andr\'e Sopczak\authorrefmark{1}, Khaled
Bekhouche\authorrefmark{5}, Chris Bowdery\authorrefmark{1}, Chris
Damerell\authorrefmark{3}, Gavin Davies\authorrefmark{1}, Lakhdar
Dehimi\authorrefmark{5}, Tim Greenshaw\authorrefmark{4}, Michal
Koziel\authorrefmark{1},  Konstantin Stefanov\authorrefmark{3}, Tim
Woolliscroft\authorrefmark{4}, Steve Worm\authorrefmark{3}
\\}
\authorblockA{\authorrefmark{1}Lancaster University, U.K.\\}
\authorblockA{\authorrefmark{5}LMSM Laboratory Biskra University,
Algeria\\}
\authorblockA{\authorrefmark{3}CCLRC Rutherford Appleton Laboratory (RAL),
U.K.\\}
\authorblockA{\authorrefmark{4}Liverpool University, U.K.\\}}

\maketitle

\begin{abstract}
Results of detailed simulations of the charge transfer inefficiency
of a prototype CCD chip are reported. The effect of radiation damage
in a particle detector operating at a future accelerator is studied
by examining two electron trap levels, 0.17\,eV and 0.44\,eV below
the bottom of the conduction band. Good agreement is found between
simulations using the ISE-TCAD DESSIS program and an analytical
model for the 0.17\,eV level. Optimum operation is predicted to be
at about 250\,K where the effect of the traps is minimal which is
approximately independent of readout frequency. This work has been
carried out within
 the Linear Collider Flavour Identification
(LCFI) collaboration  in the context of the International Linear
Collider (ILC) project.
\end{abstract}

\section{Introduction}
Particle physicists worldwide are working on the design of a high
energy collider of electrons and positrons (the International Linear
Collider or ILC) which could be operational sometime after 2016. Any
experiment exploiting the ILC will require a high performance vertex
detector to detect and measure short-lived particles. One candidate
for such a device would consist of a set of concentric cylinders of
charge-coupled devices (CCDs).

An important requirement of a vertex detector is to remain tolerant
to radiation damage for its anticipated lifetime.

CCDs suffer from both surface and bulk radiation damage. However,
when considering charge transfer losses in buried channel devices
only bulk traps are important. These defects create energy levels
between the conduction and valence band, hence electrons may be
captured by these new levels. These electrons are also emitted back
to the conduction band after a certain time. For a signal packet
this may lead to a decrease in charge as it is transferred to the
output and may be quantified by its Charge Transfer Inefficiency
(CTI), where a charge of amplitude $Q_0$ transported across $m$
pixels will have a reduced charge given by
\begin{eqnarray}
Q_m=Q_0(1-{\rm CTI})^m. \label{eqn:cti}
\end{eqnarray}
The CTI value depends on many parameters, some related to the trap
characteristics such as: trap energy level, capture cross-section,
and trap concentration (density). Operating conditions also affect
the CTI as there is a strong temperature dependence on the trap
capture rate and also a variation of the CTI with the readout
frequency. Other factors are also relevant, for example the
occupancy ratio of pixels, which influences the fraction of filled
traps in the CCD transport region. Previous studies have been
reported~\cite{Janesick,Stefanov,Ursache,Brau,Sopczak}.
\begin{figure}[htp]
\includegraphics[height=6.5cm,width=\columnwidth,clip]{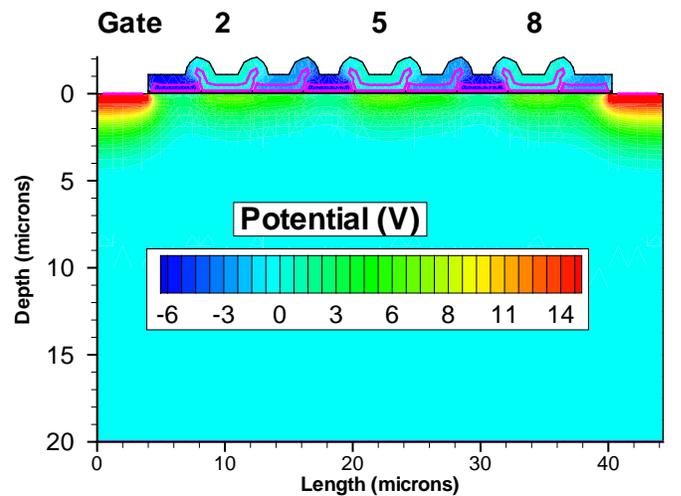}
\caption{\label{fig:trap} Detector structure and potential at gates
after initialization. The signal charge is injected under gate 2.
There are three gates for each pixel.}
\end{figure}

\section{SIMULATIONS}

The UK Linear Collider Flavour Identification (LCFI)
collaboration~\cite{LCFI,Greenshaw} has been studying a device
produced by e2V Technologies, with a manufacturer's designation
`CCD58'. It is a 2.1\,Mpixel, three-phase buried-channel CCD with
12\,$\mu$m square pixels.

Simulations of a simplified model of this have been performed with
the ISE-TCAD package (version 7.5), particularly the DESSIS program
(Device Simulation for Smart Integrated Systems). It contains an
input gate, an output gate, a substrate gate and nine further gates
(numbered 1 to 9) which form the pixels. Each pixel consists of 3
gates but only one pixel is important for this study---gates 5, 6
and 7. The simulation is essentially two dimensional but internally
there is a nominal 1\,$\mu$m device thickness (width). This is
equivalent to a thin slice of the device with rectangular pixels
12\,$\mu$m long by 1\,$\mu$m wide. The overall length and depth of
the simulated device are 44\,$\mu$m and 20\,$\mu$m respectively
(Fig.~\ref{fig:trap}).

\begin{figure}[htp]
\vspace*{-0.3cm}
\includegraphics[width=\columnwidth,height=4.5cm,clip]{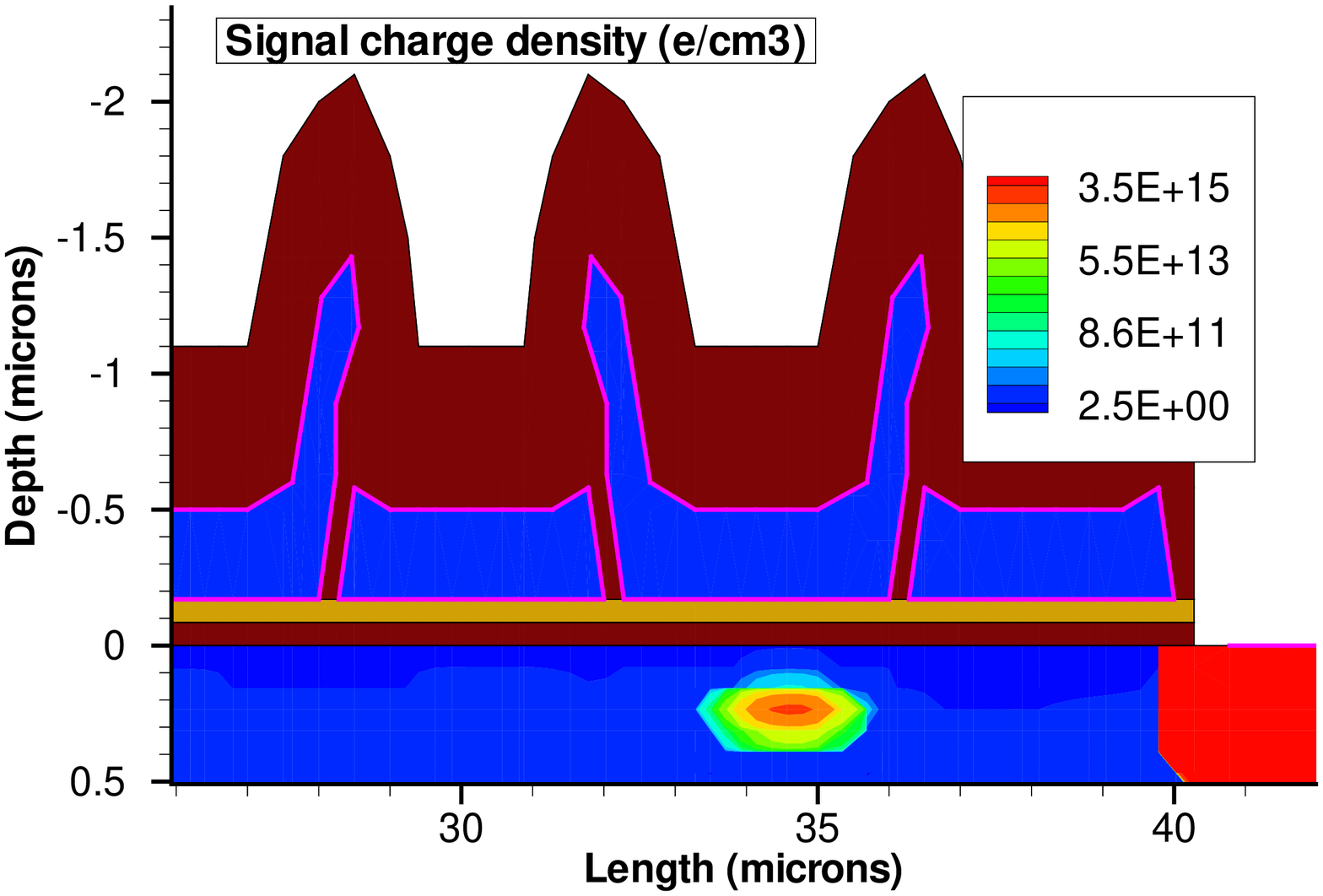}
\includegraphics[width=\columnwidth,height=4.5cm,clip]{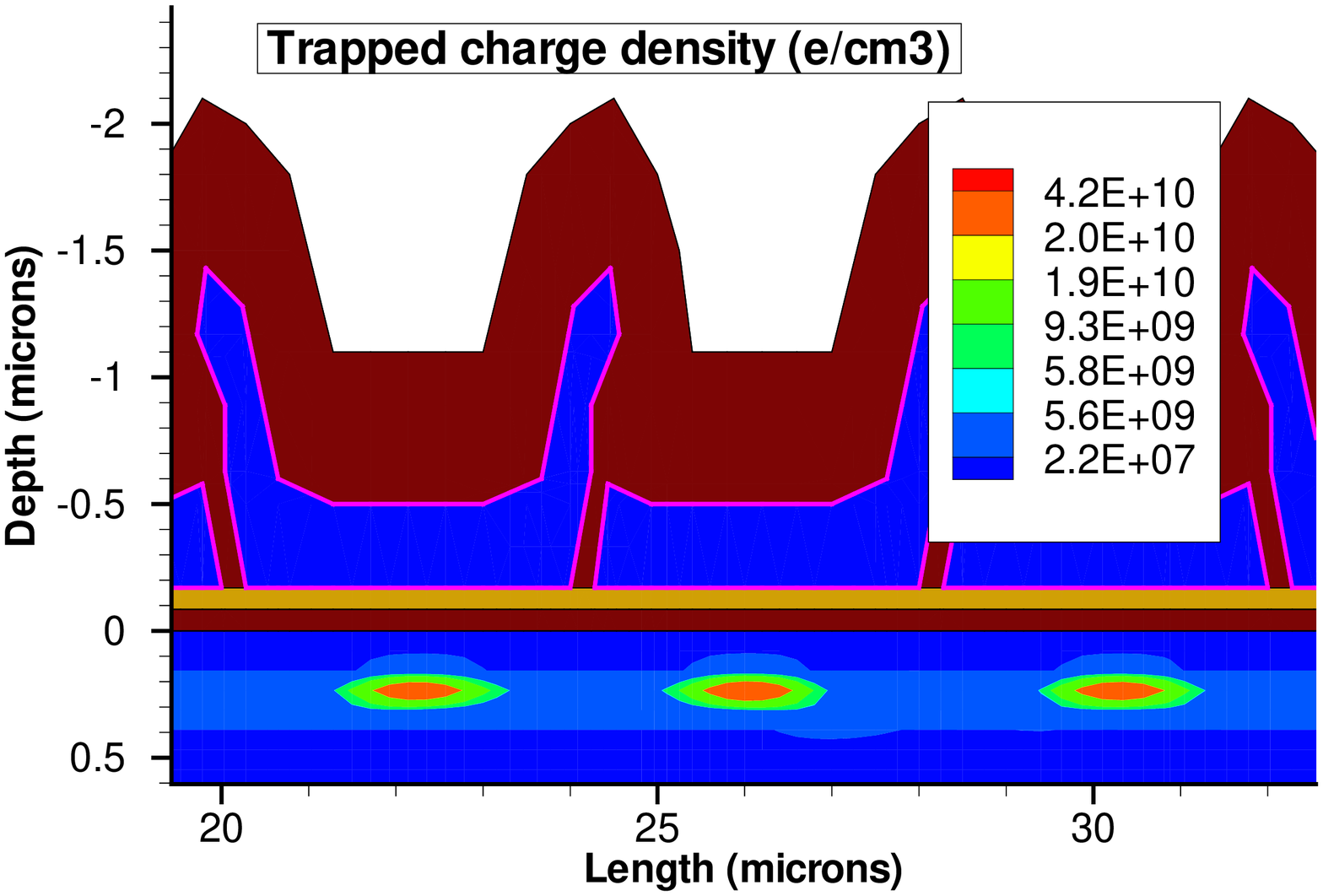}
\vspace*{-0.6cm} \caption{\label{fig:transport} Upper: Signal charge
density, almost at output gate. Lower: Trapped charge density, from
transfer of signal charge. The legend box refers to the region with
positive depth values. The thin brown line is an oxide layer and the
thin yellow line is a nitride layer.}
\end{figure}

Parameters of interest are the readout frequency, up to 50\,MHz, and
the operating temperature between 120\,K and 300\,K although
simulations have been done up to 500\,K. The charge in transfer and
the trapped charge are shown in Fig.~\ref{fig:transport}.

The signal charge used in the simulation is chosen to be similar to
the charge generated by a minimum ionising particle (MIP), amounting
to about 1620 electron-hole pairs\footnote{This number has to be
divided by 12 because the charge is assumed to be distributed over
the whole pixel but the model has only 1/12th of the true pixel
volume.} for CCD58. DESSIS has a directive for generating heavy ions
and this is exploited to create the charges. The heavy ion is made
to travel in a downwards direction starting at 1.2\,$\mu$m below
gate 2 at 1\,$\mu$s before charge transfer begins. This provides
ample time for the electrons to be drawn upwards to the transport
channel which is 0.25\,$\mu$m beneath the gate electrodes.

\subsection{Calculating CTI}
Charge Transfer Inefficiency is a measure of the fractional loss of
charge from a signal packet as it is transferred over a pixel, or
three gates. After DESSIS has simulated the transfer process, a 2D
integration of the trapped charge density distribution is performed
independently to give a total charge under each gate.

The CTI for transfer over one gate is equivalent to
\begin{eqnarray}
CTI=\frac{e_{T}-e_{B}}{e_{S}} \label{eqn:gatecti}
\end{eqnarray}
where:
\begin{itemize}
\item $e_{S}$ = electron signal packet density,
\item $e_{B}$ = background trapped electron charge density prior to signal packet transfer,
\item $e_{T}$ = trapped electron charge density under the gate, after signal transfer across gate.
\end{itemize}
 In this way the CTI is normalised for each gate.
The determinations of the trapped charge take place for gate $n$
when the charge packet just arrives at gate $n+1$. If the
determination were made only when the packet has cleared all three
gates of the pixel, trapped charge may have leaked out of the traps.

The total CTI (per pixel) is determined from gates 5, 6 and 7, hence
\begin{eqnarray}
CTI=\sum_{n=5}^7 \frac{e_{T}-e_{B}}{e_{S}} \label{eqn:pixelcti}
\end{eqnarray}
where $n$ is the gate number.  The background charge is taken as the
trapped charge under gate 2 because this gate is unaffected by the
signal transport when the charge has just passed gates being
processed.

\subsection{Initial tests of the DESSIS program}

DESSIS simulations have been carried out with a zero concentration
of electron traps as in unirradiated silicon.

The DESSIS program is steered with a command file which contains
electrode voltage values for each of the three phases as a function
of time. For these simulations the voltage values (peak value 7\,V),
which were originally digitised from real experiments with CCD58,
were replaced by reduced values without altering the frequency or
phase.

Since there were no traps there was no trapped charge so a different
estimator of CTI is required. The electron charge density left under
gate $n$ when the charge packet has moved to gate $n+1$ gives the
partial CTI estimator. It is normalised by the original electron
charge density. As before, the CTI for a pixel is computed by adding
the partial CTI's for gates 5, 6, and 7.

\begin{figure}[Htp]
\includegraphics[height=6.2cm,width=\columnwidth,clip]{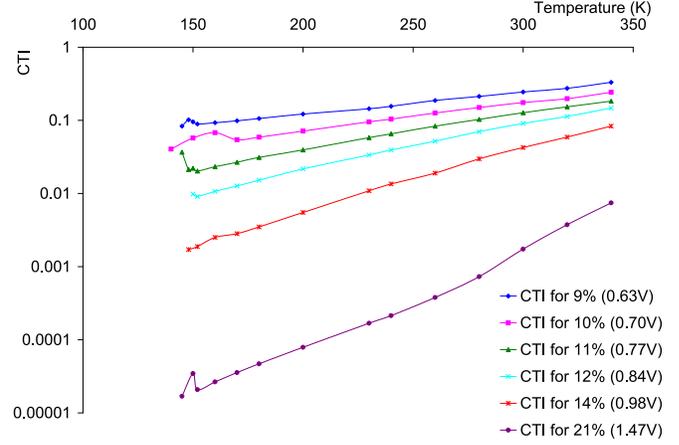}
\caption{\label{fig:notraps1}CTI values against temperature for
simulations with no traps at a clocking frequency of 50\,MHz.}
\end{figure}

\begin{figure}[Htpb]
\includegraphics[height=5.2cm,width=\columnwidth,clip]{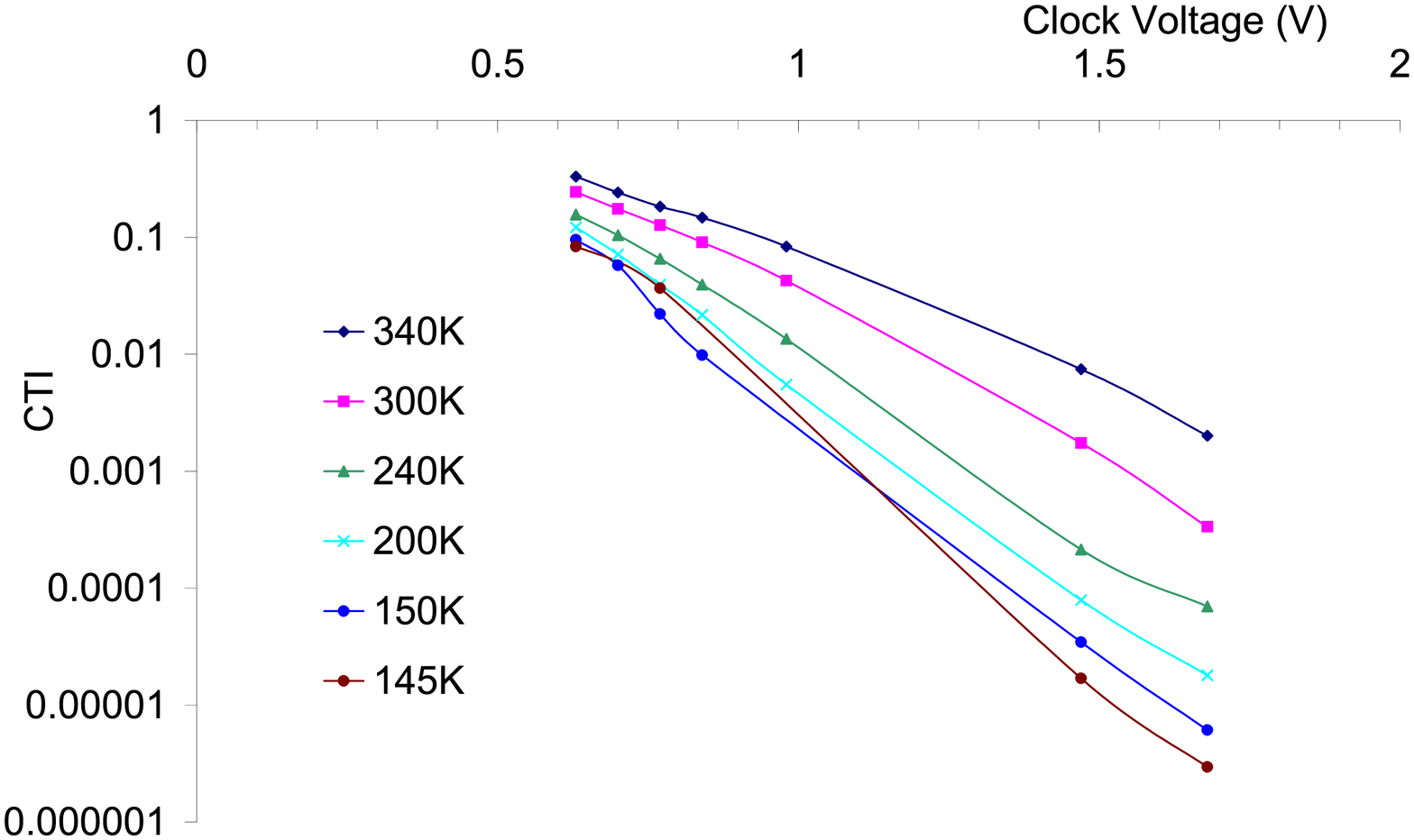}
\vspace*{-0.5cm}\caption{\label{fig:notraps2}CTI values against
clock voltage for simulations with no traps at a clocking frequency
of 50\,MHz.}
\end{figure}

Figure~\ref{fig:notraps1} shows the variation of CTI with
temperature for various clock voltages. Figure~\ref{fig:notraps2}
shows the variation of CTI with clock voltage for a range of
temperatures. The device operates with a negligible CTI above 3\,V
and has a large CTI below 1\,V. Also the CTI grows with temperature.

Examination of snapshots of electron density plots produced by
DESSIS during a simulation run confirm that electrons leak out of
the main charge packet during transfer at reduced clock voltages
leading to an even distribution of electrons under all of the gates.

\subsection{0.17\,eV and 0.44\,eV traps} This CTI study, at nominal clock voltage, focuses only on
the bulk traps with energies 0.17\,eV and 0.44\,eV below the bottom
of the conduction band. These will be referred to simply as the
0.17\,eV and 0.44\,eV traps. An incident particle with sufficient
energy is able to displace an atom from its lattice point leading
eventually to a stable defect. These defects manifest themselves as
energy levels between the conduction and valence band, in this case
the energy levels 0.17\,eV and 0.44\,eV; hence electrons and/or
holes may be captured by these levels. The 0.17\,eV trap is an
oxygen vacancy defect, referred to as an A-centre defect.
The 0.44\,eV trap is a phosphorus-vacancy defect---an E-centre
defect---that is, a result of the silicon being doped with
phosphorus and a vacancy manifesting from the displacement of a
silicon atom bonded with the phosphorus atom \cite{Stefanov}.

In order to determine the trap densities for use in simulations, a
literature search on possible ILC radiation backgrounds and trap
induction rates in silicon was undertaken. The main expected
background arises from e$^+$e$^-$ pairs with an average energy of
10\,MeV and from neutrons (knocked out of nuclei by synchrotron
radiation).

Table~1 shows results of background simulations of e$^+$e$^-$ pairs
generation for three proposed vertex detector designs (from three
ILC detector concepts).
\begin{center}
\begin{tabular}{|c | c| c| c|}
\hline Simulator & SiD & LDC & GLD \\
\hline CAIN/Jupiter & 2.9 & 3.5 & 0.5 \\
GuineaPig & 2.3 & 3.0 & 2.0 \\
\hline
\end{tabular}
\end{center}
{\footnotesize\rm Table 1. Simulated background results for three
different detector scenarios. The values are hits per square
centimetre per e$^+$e$^-$ bunch crossing. SiD is the Silicon
Detector Concept~\cite{SiD}, LDC is the Large Detector
Concept~\cite{LDC} and GLD is the Global Linear collider
Detector~\cite{GLD}. }

Choosing the scenario with the highest expected background, that is
the LDC concept, where the innermost layer of the vertex detector
would be located 14\,mm from the interaction point, one can estimate
an e$^+$e$^-$ flux around 3.5\,hits/cm$^2$/bunch crossing which
gives a fluence of 0.5$\times 10^{12}$\,e/cm$^2$/year. In the case
of neutrons, from two independent studies, the fluence was estimated
to be 10$^{10}$\,n/cm$^2$/year~\cite{Maruyama} and 1.6$\times
10^{10}$\,n/cm$^2$/year~\cite{Vogel}.

Based on the literature~\cite{Marconi, robbins, robbins-roy,
walker,wertheim,suezawa, saks, srour,fretwurst}, the trap densities
introduced by 1\,MeV neutrons and 10\,MeV electrons have been
estimated with two established assumptions: the electron trap
density is a linear function of dose, and the dose is a linear
function of fluence. A summary is given in Table~2.

\vspace*{0.2cm}
\begin{center}
\begin{tabular}{| l | l| l |}\hline

 Particle type & 0.17\,eV (\cm{-3}) & 0.44\,eV (\cm{-3})  \\ \hline
 10 MeV e$^-$&   \ntp{3.0}{11} & \ntp{3.0}{10}\\
 \hphantom{0}1 MeV n &   \ntp{(4.5\ldots 7.1)}{8} & \ntp{(0.7\ldots 1.1)}{10}
 \\ \hline
 total    &   \ntp{3.0}{11} & \ntp{4.1}{10}\\ \hline
\end{tabular}
\end{center}
 {\footnotesize\rm Table 2. Estimated densities of
traps after irradiation for one year. For neutrons, the literature
provides two values.}
 \vspace*{0.2cm}

\noindent The actual trap concentrations and electron capture
cross-sections used in the simulations are shown in Table~3.

\vspace*{0.5cm}
\begin{center}
\begin{tabular}{| c | l| l| l |}\hline
$\asb{E}{t}-\asb{E}{c}$ (eV)
 & Type & $C$ (\cm{-3}) &$\sigma$ (\cm{2})\\ \hline
0.17 & Acceptor & \ntp{1}{11} & \ntp{1}{-14}\\
0.44 & Acceptor & \ntp{1}{11} & \ntp{3}{-15}\\ \hline
\end{tabular}
\end{center}
 {\footnotesize\rm Table 3. Trap concentrations
(densities) and electron capture cross-sections as used in the
DESSIS simulations.}

\subsection{Partially Filled Traps}
Each electron trap in the semiconductor material can either be {\em
empty\/} (holding no electron) or {\em full\/} (holding one
electron). In order to simulate the normal operating conditions of
CCD58, partial trap filling was employed in the simulation (which
means that some traps are full and some are empty) because the
device will transfer many charge packets during continuous
operation.

In order to reflect this, even though only the transfer of a single
charge packet was simulated, the following procedure was followed in
all cases. From t\,=\,0 seconds to t\,=\,98\,$\mu$s, the gates ramp
up and are biased in such a way to drain the charge to the output
drain. The device is in a fully normal biased state then at
98\,$\mu$s.  To obtain partial trap filling, the simulation
waits\footnote{This waiting time is calculated from a 1\% mean pixel
occupancy with a 50\,MHz readout frequency.} 2\,$\mu$s between
98\,$\mu$s and 100\,$\mu$s to allow traps to partially empty. The
test charge is generated at 99\,$\mu$s. The simulation then starts
the three clock phases, varying voltage with time to cause the
transfer of the signal charge packet through the device.

\section{Analytical Models}
The motivation for introducing the following two simple analytical
models is to understand the underlying effects and to make
comparisons with the DESSIS simulations (referred to as the ``full
simulations'').

\subsection{Simple CTI Model}
Firstly, a simple analytical model is considered, based upon a
single trapping level---a so-called Simple CTI model. This is
significantly faster than a full simulation. It also provides a
simple method to see the effect of changing parameters and
demonstrates physics understanding.

The charge transfer process is modelled by a differential equation
in terms of the different time constants and temperature dependence
of the electron capture and emission processes. In the electron
capture process, electrons are captured from the signal packet and
each captured electron fills a trap. This occurs at the capture rate
$\tau_{\rm c}$. The electron emission process is described by the
emission of captured electrons from filled traps back to the
conduction band, and into a second signal packet at the emission
rate $\tau_{\rm e}$.

\subsubsection{Capture and emission time constants}
The Shockley-Read-Hall theory~\cite{ShockleyReadHall} considers a
defect at an energy $E_{\rm t}$ below the bottom of the conduction
band, $E_{\rm c}$, and gives
\begin{eqnarray}
\tau_{\rm c} &=& \frac{1}{\sigma_{\rm e} \nu_{\rm th} n_{\rm s}} \label{eqn:capture}\\
\tau_{\rm e} &=& \frac{1}{\sigma_{\rm e} \chi_{\rm e} \nu_{\rm th}
N_{\rm c}}\exp{\left(\frac{E_{\rm c} - E_{\rm t}}{k_{\rm
B}T}\right)} \label{eqn:emission}
\end{eqnarray}
where:
\begin{itemize}
\item $\sigma_{\rm e}$ = electron capture cross-section
\item $\chi_{\rm e}$ = entropy change factor by electron emission
\item $\nu_{\rm th}$ = electron thermal velocity
\item $N_{\rm c}$ = density of states in the conduction band
\item $k_{\rm B}$ = Boltzmann's constant
\item $T$ = absolute temperature
\item $n_{\rm s}$ = density of signal charge packet.
\end{itemize}
It is assumed that $\chi_{\rm e}=1$.

At low temperatures, the emission time constant $\tau_{\rm e}$ can
be very large and of the order of seconds. The charge shift time is
of the order of nanoseconds. A larger $\tau_{\rm e}$ means that a
trap remains filled for much longer than the charge shift time.
Further trapping of signal electrons is not possible and,
consequently, CTI is small at low temperatures. A peak occurs
between low and high temperatures because the CTI is also small at
high temperatures. This manifests itself because, at high
temperatures, the emission time constant decreases to become
comparable to the charge shift time. Now, trapped electrons rejoin
their signal packet.

\subsubsection{Charge Transfer Equation}
From the fraction of filled traps, the following differential
equation can be derived:
\begin{eqnarray}
\frac{\mathrm{d}r_{\rm f}(t)}{\mathrm{d}t} =\frac{1-r_{\rm
f}(t)}{\tau_{\rm c}}-\frac{r_{\rm f}(t)}{\tau_{\rm e}}
\end{eqnarray}
where $r_{\rm f}(t)$ is the time-dependent fraction of filled traps
\begin{eqnarray}
r_{\rm f}(t)=\frac{n_{\rm t}(t)}{N_{\rm t}}
\end{eqnarray}
\begin{itemize}
\item $n_{\rm t}(t)$ = density of traps filled by electrons
\item $N_{\rm t}$ = density of traps
\end{itemize}
Considering that the traps are partially filled and using the
initial condition:
\begin{eqnarray}
r_{\rm f}(0) = r_{\rm f}(t_{\rm sh})e^{-t_{\rm w}/\tau_{\rm e}}
\label{eqn:initial}
\end{eqnarray}
where $r_{\rm f}(0)$ is the fraction of filled traps after a mean
waiting time, $t_{\rm w}$, the differential equation can be solved
to provide an expression for the CTI:
\begin{eqnarray}
CTI = \frac{3N_{\rm t}}{n_{\rm s}}\left(r_{\rm f}(t)-r_{\rm
f}(0)\right) \label{eqn:cti2}
\end{eqnarray}
\begin{eqnarray}
r_{\rm f}(t) = (r_{\rm f}(0)-\tau_{\rm s}/\tau_{\rm
c})e^{-t/\tau_{\rm s}+\tau_{\rm s}/\tau_{\rm c}} ,
\frac{1}{\tau_{\rm s}}=\frac{1}{\tau_{\rm c}}+\frac{1}{\tau_{\rm e}}
\label{eqn:initial2}
\end{eqnarray}
\begin{eqnarray}
CTI=\frac{3N_{\rm t}}{n_{\rm s}}\left(\frac{\tau_{\rm s}}{\tau_{\rm
c}}-r_{\rm f}(0)\right)\left(1-e^{-t_{\rm sh}/\tau_{\rm s}}\right)
\label{eqn:cti3}
\end{eqnarray}
where $t_{\rm sh}$ is the shift-time. For one gate, $t_{\rm sh}=1/(3{\rm f})$, where
${\rm f}$ is the readout frequency.

This definition is for CTI for a single trap level. The factor of
three appears since there is a sum over the three gates that make up
a pixel.

\subsubsection{Matching the CTI definition of the
simulation}

The Simple Model has been adapted by including initially filled
traps and by the incorporation of a so-called $P$ factor to CTI:
\begin{equation}
P = e^{-t_{\rm sh}/\tau_{\rm e}} + e^{-2t_{\rm sh}/\tau_{\rm e}} +
e^{-3t_{\rm sh}/\tau_{\rm e}}
\end{equation}
This models the situation where the trapped charge under gate 5
started to empty at time $t$ minus three shift-times, that under
gate 6 at $t$ minus two shift-times  and that under gate 7 at $t$
minus one shift-time. An alternative factor, called $P^\prime$, has
also been used to compare with simulated data. This is defined as:
\begin{equation}
P^\prime = 1 + e^{-t_{\rm sh}/\tau_{\rm e}} + e^{-2t_{\rm
sh}/\tau_{\rm e}}
\end{equation}
and models the situation one shift-time earlier than for $P$.

\subsection{Improved Model}
     The second analytical model that has been developed is referred
to as the Improved Model (IM), based on the work of T.~Hardy et
al.~\cite{hardy}. It is improved by adjusting initial assumptions to
fit the study of CCD58. The Improved model also considers the effect
of a single trapping level, but only includes the emission time in
its differential equation:
\begin{eqnarray}
\frac{\mathrm{d}n_{\rm t}}{\mathrm{d}t}=-\frac{n_{\rm t}}{\tau_{\rm
e}}
\end{eqnarray}
where $n_{\rm t}$ is the density of filled traps. The traps are
initially filled for this model and $\tau_{\rm c}\ll t_{\rm sh}$.
Nevertheless, to be consistent with the full DESSIS simulations
(that use partially filled traps) the Improved Model uses a time
constant between the filling of the traps such that the traps remain
partially filled when the new electron packet passes through the
CCD. The solution of this differential equation leads to another
estimator of the CTI:

\begin{eqnarray}
CTI^{I}=\left(1-e^{-t_{\rm sh}/\tau_{\rm c}}\right)\frac{3N_{\rm
t}}{n_{\rm s}}\left(e^{-t_{\rm join}/\tau_{\rm e}}-e^{-t_{\rm
emit}/\tau_{\rm e}}\right) \label{eqn:cti5}
\end{eqnarray}
\begin{itemize}
\item $t_{\rm emit}=t_{\rm w}$ is the total emission time from the
previous packet.
\item $t_{\rm join}$ is the time period during which the charges can join
the parent charge packet.
\end{itemize}

\section{Simulation Results}

The CTI dependence on temperature and readout frequency was explored
using ISE-TCAD simulations.

\subsubsection{0.17\,eV traps}

\begin{figure}[htp]
\includegraphics[height=6.6cm,width=\columnwidth,clip]{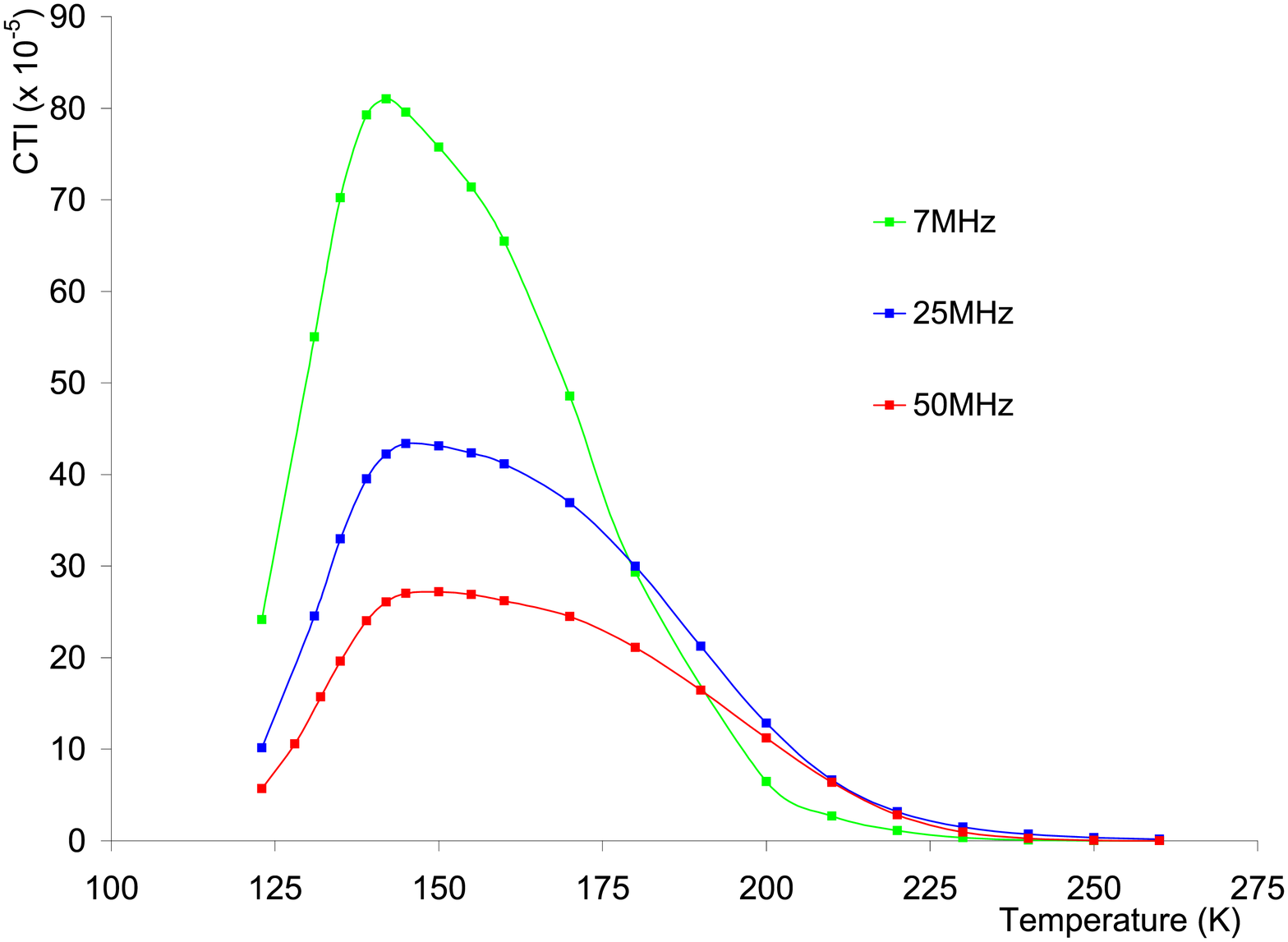}
\caption{\label{fig:allfreq_0_17eV}CTI values for simulations with
0.17\,eV partially filled traps at clocking frequencies 7, 25 and
50\,MHz.}
\end{figure}

Figure~\ref{fig:allfreq_0_17eV} shows the CTI for simulations with
partially filled 0.17\,eV traps at different frequencies for
temperatures between 123\,K and 260\,K, with a nominal clock voltage
of 7\,V.

\begin{figure}[htpb]
\includegraphics[height=5.75cm,width=\columnwidth,clip]{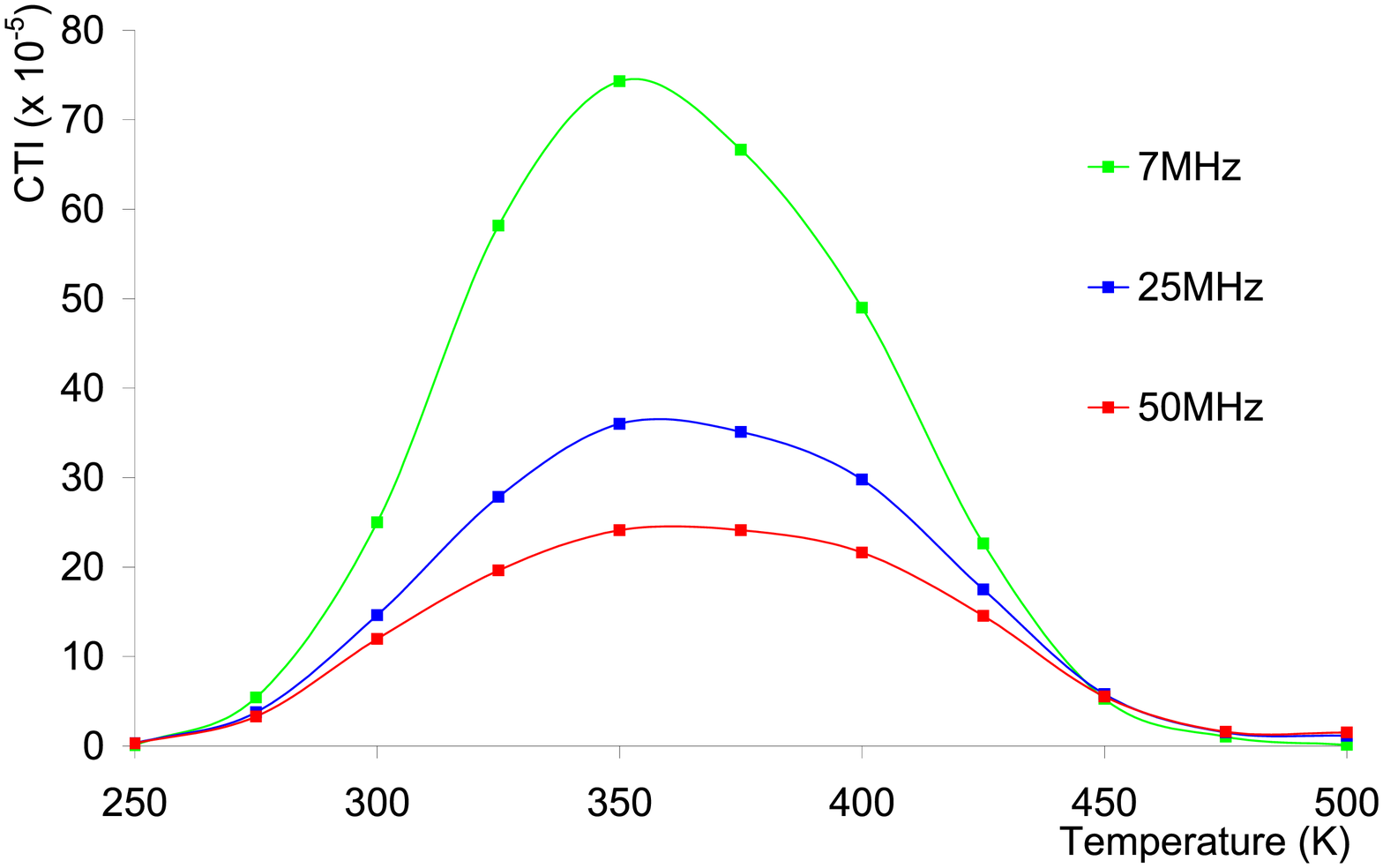}
\caption{\label{fig:allfreq_0_44eV}CTI values for simulations with
0.44\,eV partially filled traps at clocking frequencies 7, 25 and
50\,MHz.}
\end{figure}

A peak structure can be seen. For 50\,MHz, the peak is at 150\,K
with a CTI of $27.2 \times 10^{-5}$. The peak CTI is in the region
between 145\,K and 150\,K for a 25\,MHz clock frequency and with a
value of about $43 \times10^{-5}$. This is about 1.6 times bigger
than the charge transfer inefficiency at 50\,MHz. The peak CTI for
7\,MHz occurs at about 142\,K, with a maximum value of about $81
\times 10^{-5}$, an increase from the peak CTI at 50\,MHz $(27
\times 10^{-5})$ by a factor of about 3 and an increase from the
peak CTI at 25\,MHz $(43 \times 10^{-5})$ by a factor of nearly 2.
Thus CTI increases as frequency decreases. For higher readout
frequency there is less time to trap the charge, thus the CTI is
reduced. At high temperatures the emission time is so short that
trapped charges rejoin the passing signal.

\subsubsection{0.44\,eV traps}

Simulations were also carried out with partially filled 0.44\,eV
traps at temperatures ranging from 250\,K to 500\,K. This is because
previous studies~\cite{Sopczak} on 0.44\,eV traps have shown that
these traps cause only a negligible CTI at temperatures lower than
250\,K due to the long emission time and thus traps remain fully
filled at lower temperatures. The results are depicted in
Fig.~\ref{fig:allfreq_0_44eV}.

The peak CTI is higher for lower frequencies with little temperature
dependence of the peak position.

\subsubsection{0.17\,eV and 0.44\,eV traps together}

\begin{figure}[htp]
\includegraphics*[height=6.3cm,width=\columnwidth,clip]{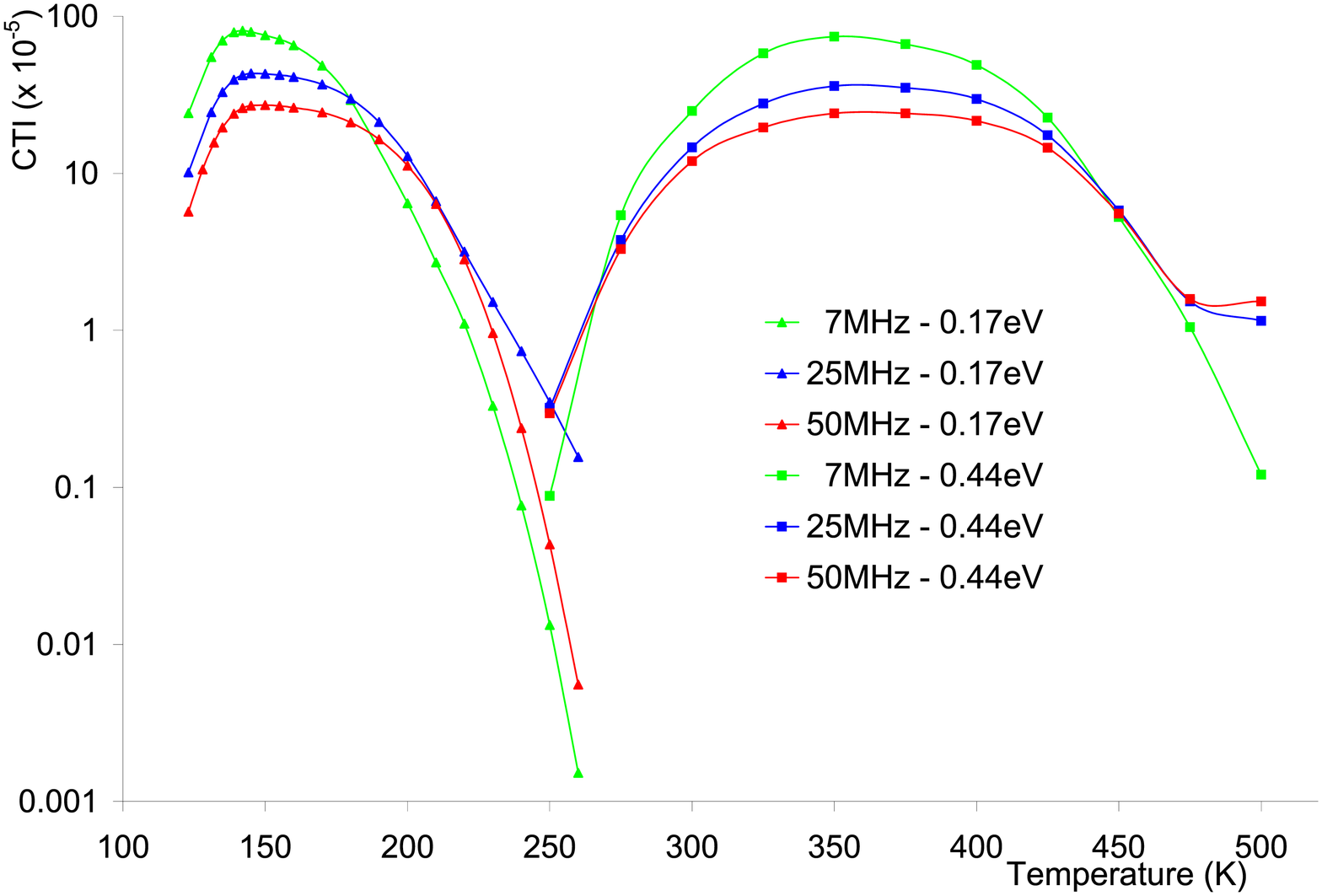}
\caption{\label{fig:all_log}CTI values for simulations for partially
filled 0.17\,eV and 0.44\,eV traps. Comparison of CTI at frequencies
7, 25 and 50\,MHz for different trap energy level on a logarithmic
scale.}
\end{figure}

The logarithmic scale view (Fig. \ref{fig:all_log}) of the
simulation results at the different frequencies and trap energies
clearly identifies an optimal operating temperature of about 250\,K.

\begin{figure}[htp]
\includegraphics*[height=7.0cm,width=\columnwidth,clip]{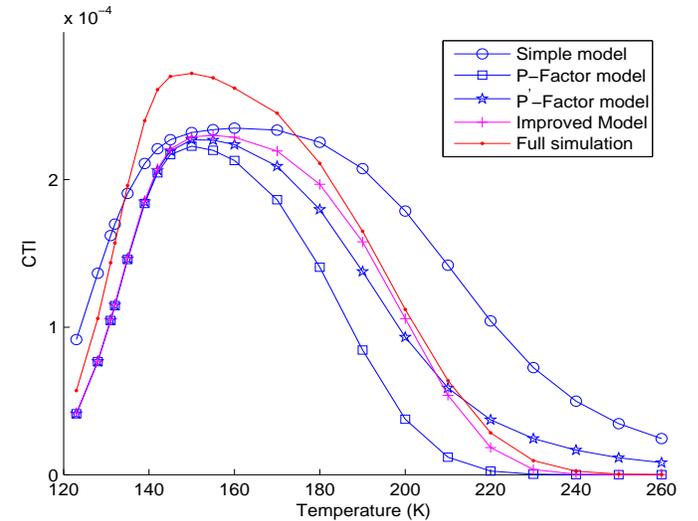}
\caption{\label{fig:diff_models}CTI values against temperature for
different models with 50\,MHz readout. See text for details of the
Simple Model and its adaption with the $P$ and $P^\prime$ factors
and the Improved Model.}
\end{figure}

\subsection{Comparisons with Models}

Figure~\ref{fig:diff_models} shows that the basic Simple Model does
not agree well with the full simulation. Applying the $P$ factor
appears to overcompensate for the deficiencies and the $P^\prime$
factor gives a reasonable but not perfect agreement.

\begin{figure}[htp]
\includegraphics[height=5.75cm,width=\columnwidth,clip]{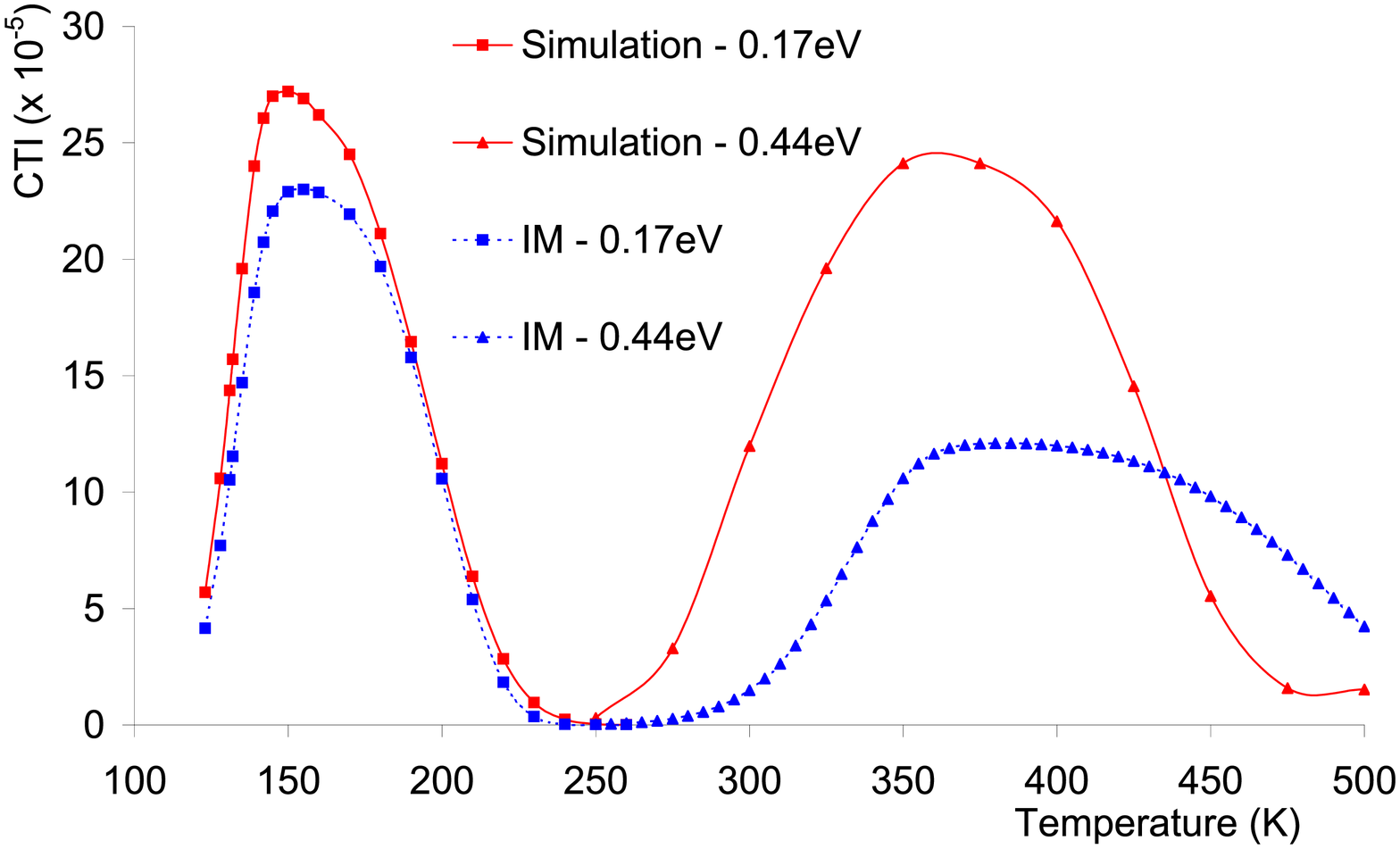}
\vspace*{2mm} \caption{\label{fig:ISE_v_IM}CTI values for
simulations for 0.17\,eV versus 0.44\,eV partially filled traps at
clocking frequency 50\,MHz. Comparison of Improved Model (IM) with
full DESSIS simulation.}
\end{figure}

Figure~\ref{fig:ISE_v_IM} compares the full DESSIS simulation for
0.17\,eV and 0.44\,eV traps and clocking frequency of 50\,MHz to the
Improved Model. It emphasises the good agreement between the model
and full simulations at temperatures lower than 250\,K with 0.17\,eV
traps, but shows a disagreement at higher temperatures for the
0.44\,eV traps.

\begin{figure}[htp]
\includegraphics*[height=6.6cm,width=\columnwidth,clip]{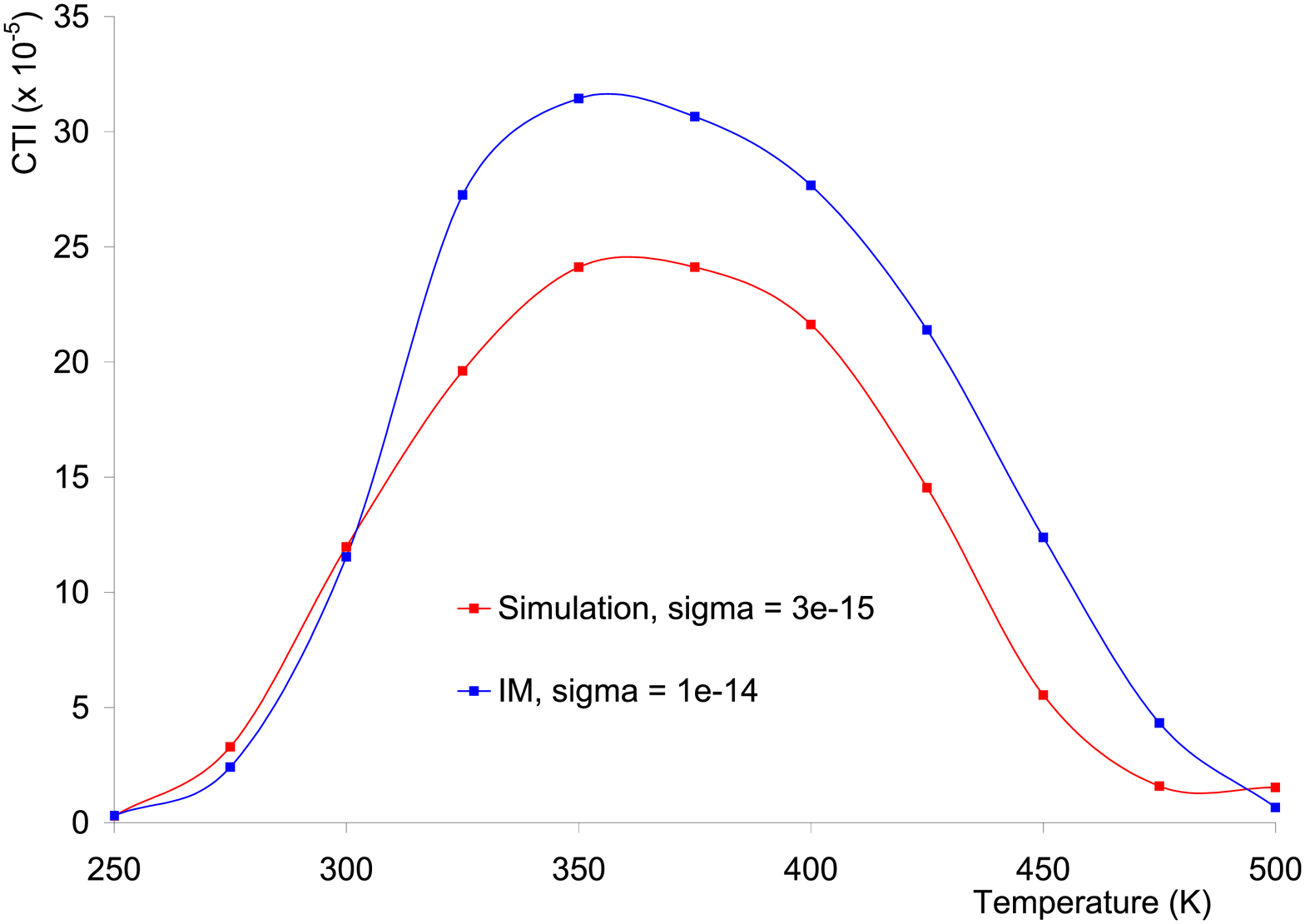}
\vspace*{2mm} \caption{\label{fig:ISEvIM2}CTI values for simulations
with 0.44\,eV partially filled traps at clocking frequency 50\,MHz.
Comparison of Improved Model (IM) ($\sigma= 10^{-14}$\,cm$^2$) with
full DESSIS simulation ($\sigma= 3 \times 10^{-15}$\,cm$^2$).}
\end{figure}

If the 0.44\,eV trap electron capture cross-section in the Improved
Model is increased to $10^{-14}$\,cm$^{2}$, a somewhat better
agreement is found, as shown in Figure~\ref{fig:ISEvIM2}. However it
is clear that there are limitations with the Improved Model. They
could relate to a breakdown of the assumptions at high temperatures,
to ignoring the precise form of the clock voltage waveform, or to
ignoring the pixel edge effects. Further studies are required.

\section{Conclusions}

The Charge Transfer Inefficiency (CTI) of a CCD device has been
studied with a full simulation (ISE-TCAD DESSIS) and compared with
analytical models.

Partially filled traps from the 0.17\,eV and 0.44\,eV trap levels
have been implemented in the full simulation and variations of the
CTI with respect to temperature and frequency have been analysed.
The results confirm the dependence of CTI with the readout
frequency. At low temperatures ($<250$\,K) the 0.17\,eV traps
dominate the CTI, whereas the 0.44\,eV traps dominate at higher
temperatures.

A large emission time constant $\tau_{\rm e}$ results in a trap
remaining filled for much longer than the charge shift time. Further
trapping of signal electrons is not possible so the CTI is small at
low temperatures. At high temperatures the emission time constant
decreases to become comparable to the charge shift time. Trapped
electrons rejoin their signal packet and, because most are emitted
during the charge transfer time, there is again a small CTI. For
intermediate temperatures, a clear peak structure is observed.

Good agreement between simulations and a so-called Improved Model
has been found for 0.17\,eV traps but not for 0.44\,eV traps. This
shows the limitations of the Improved Model with respect to the full
simulation.

The optimum operating temperature for CCD58 in a high radiation
environment is found to be about 250\,K.

Interest is now moving to alternative CCD designs, particularly
2-phase column-parallel readout devices. The extensive amount of
research that has been carried out on CCD58 contributes to the
development of future CCD designs.

\section*{Acknowledgments}
This work is supported by the Particle Physics and Astronomy
Research Council (PPARC) and Lancaster University. The Lancaster
authors wish to thank Alex Chilingarov, for helpful discussions, and
the particle physics group at Liverpool University, for the use of
its computers.

\end{document}